\documentstyle[12pt]{article}

\renewcommand{\theequation}{\arabic{section}.\arabic{equation}}

\renewcommand{\thesection}{\arabic{section})}

 \newcommand{\be}{\begin{equation}}
 \newcommand{\ee}{\end{equation}}
 \newcommand{\ba}{\begin{eqnarray}}
 \newcommand{\ea}{\end{eqnarray}}
 
 \newcommand{\del}{\partial}

\def\B{\tilde B}

\def\C{\tilde C}

\newcommand{\expo}{\exp \lef \{ - \int d^3z }

\newcommand{\lef}{\left}

\newcommand{\ri}{\right}

\newcommand{\bb}{\bar{B}}

\newcommand{\cf}{{\cal F}}

\newcommand{\cl}{{\cal L}}

\newcommand{\fr}{\frac}

\begin{document}

\begin{titlepage}

\topmargin -15mm

\rightline{\bf UFRJ-IF-FPC-006/96}

\vskip 10mm

\centerline{ \LARGE\bf Local Charged States of the Gauge Field}
\vskip 2mm
\centerline{ \LARGE\bf in Three Dimensional Maxwell-Type Theories}

    \vskip 2.0cm

    \centerline{\sc E.C.Marino }

     \vskip 0.6cm
     
    \centerline{\it Instituto de F\'\i sica }
    \centerline{\it Universidade Federal do Rio de Janeiro }
    \centerline{\it Cx.P. 68528}
    \centerline{\it Rio de Janeiro RJ 21945-970}
    \centerline{\it Brasil}

\vskip 1.0cm

\begin{abstract} 
 
Gauge invariant local creation operators of charged states are 
introduced and studied in pure gauge theories of the Maxwell type in
2+1D. These states are usually unphysical because of the subsidiary 
condition imposed on the physical subspace by Gauss' law. A dual Maxwell
theory which possesses a topological electric charge is introduced. 
Pure Electrodynamics lies in the sector where the topological charge 
identically vanishes. Charge bearing operators fully expressed in terms of
the gauge field, however, can create physical 
states in the nontrivial topological
sectors which thereby generalize QED. An order disorder structure exists
relating the charged operators and the
magnetic flux creating (vortex) operators, both through commutation rules 
and correlation functions. The relevance of this structure for bosonization
in 2+1D is discussed.

\end{abstract}

\vskip 3cm
$^*$ Work supported in part by CNPq-Brazilian National Research Council.
     E-Mail address: marino@if.ufrj.br

\end{titlepage}

\hoffset= -10mm

\leftmargin 23mm

\topmargin -8mm
\hsize 153mm
 
\baselineskip 7mm
\setcounter{page}{2}

\section{Introduction}

A rather interesting problem in Quantum Field Theory is the obtainement of 
the operators creating the quantum states carrying the topological charge
of a certain system. In three dimensional space, the topological charge is
the magnetic flux along the spatial plane and this problem has been solved
years ago \cite{pol,vor,nv,rede,cont}. The underlying framework for 
this is an abelian gauge theory in 2+1D. Given these 
topological charge creating operators (vortex operators), 
there is usually an order-disorder duality 
structure relating them to dual operators
carrying another conserved charge which in 2+1 dimensions is the electric
charge. The construction of this second type of operators is also an 
appealing problem. Moreover, it is known that this type of order-disorder
structure is responsible for bosonization in 1+1D \cite{odd,cm} and also
in higher dimensions \cite{bos,mar}. 

A charge bearing operator dual to the vortex operator was introduced 
recently in Maxwell-type theories \cite{ap}. 
That construction, however, besides being rather
cumbersome, has the disadvantage of being non gauge invariant. In addition
the states created by that operators did not satisfy the subsidiary 
condition imposed by Gauss' law on the physical states. 
More recently charge bearing local gauge invariant operators were 
obtained in Chern-Simons theory, both pure 
and coupled to fermions \cite{cha}. 
In the present work
we introduce local gauge invariant charge bearing operators in 
Maxwell-type theories in 2+1D and make a detailed study of their properties
and correlation functions. The present construction 
is much simpler and transparent than that of \cite{ap}. 
We also introduce a theory which is dual to  
Maxwell QED, in which the electric charge appears as the topological charge.
Pure Electrodynamics corresponds to the sector were the toplogical charge 
is identically zero. The nontrivial sectors consist in a generalization of
QED. An analogous system was introduced recently in 3+1D \cite{qed}. In 
this theory, the charge carrying operators dual to the vortex field do
belong to the physical subspace.

In Section 2, we introduce the charge carrying $\sigma$-operators in
Maxwell theory and evaluate its commutators with the charge and with the
magnetic flux carrying vortex $\mu$-operator. The locality of its correlation
functions is also demonstrated in general. In Section 3, we evaluate the
correlation functions of $\sigma$  as well as the  mixed 
$\sigma-\mu$-correlation functions both in Maxwell theory and in a nonlocal
version of it which appeared in the bosonization of a free massless fermion
field in 2+1D \cite{bos}. This theory, given by (\ref{mnl}), was also 
shown to describe the real electromagnetic interaction of four-dimensional
charged particles constrained to move on a plane \cite{qedpl}.
In Section 4, we introduce the dual Maxwell theory and the generalization 
of it containing nontrivial topological sectors. The charge bearing operators
are introduced and their correlation functions, explicitly evaluated.
Two Appendixes are also included.

\section{ The $\sigma$ Operator}
\setcounter{equation}{0}   

\subsection{Correlation Functions}

Let us start with Maxwell theory and introduce the operator $\sigma$ which
will create the charged states of the gauge field and whose correlation
functions are going to be local. For these, inspired in the construction
of similar operators in Chern-Simons theory \cite{cha}, we write
\be
<\sigma(x) \sigma^\dagger(y)> = Z^{-1} \int DA_\mu \expo 
\lef[ \fr{1}{4}
F_{\mu\nu}F^{\mu\nu} + F_{\mu\nu} \C^{\mu\nu}
+ \fr{1}{2} \C_{\mu\nu}\C^{\mu\nu}  \ri]\ri\}
\label{cf1}
\ee
where $\C_{\mu\nu}= \C_{\mu\nu}(z;x)-\C_{\mu\nu}(z;y)$ 
with $\C_{\mu\nu}(z,x)$ given by
$$
\C_{\mu\nu}(z;x) = \del_\nu \C_\mu(z;x)
$$
with
\be
\C_\mu(z;x) = i a \int_{x,L}^\infty d\xi_\mu \lef [\fr{1}{-\Box}\ri] (z- \xi)
\label{c}
\ee
Here $[\fr{1}{-\Box}] = \fr{1}{4\pi|x|}$ is the Fourier transform of $1/k^2$.
As in previous cases \cite{cha,nv} the second 
term in (\ref{cf1}) corresponds to the operator itself and the last term is
a renormalization factor which removes unphysical parts 
and makes (\ref{cf1}) independent of the line $L$ appearing in 
the definition of the ``external field'' $\C_{\mu\nu}$ in (\ref{c}). 
This will be shown explicitly below.

\subsection{Commutation Rules}

From (\ref{cf1}) we
can extract the form of $\sigma$, namely
$$
\sigma(x) = \exp \lef \{- i \int d^3z \C_{\mu\nu}F^{\mu\nu}
 \ri \}
$$
$$
\sigma(x) = \exp \lef \{ ia \int_{x,L}^\infty d\xi_\beta \int d^3z
\lef [\fr{1}{-\Box}\ri](\xi -z) \del_\alpha F^{\alpha\beta}(z)
 \ri \} 
$$
\be
\sigma(x) = \exp \lef \{ ia \int_{x,L}^\infty d\xi^i A_i(\xi) 
+ \int d^3z
\lef [\fr{1}{-\Box}\ri](x-z) \del_\alpha A^\alpha (z)
 \ri \} 
\label{si}
\ee
We see that $\sigma$ is gauge invariant and the first part of it is
the Wilson line.

Let us consider now the commutator of 
$\sigma$ with the charge density operator
$j^0=\del_i E^i$. We can always choose the gauge in such a way that the
second part commutes with $j^0$. The first part immediately gives
\be
[j^0(x),\sigma(y)] = a \sigma \delta^2 (x-y)
\label{com1}
\ee
showing that $\sigma$ carries $a$ units of charge. The states created by
$\sigma$, however, are not physical because by virtue of (\ref{com1})
they do not satisfy the 
subsidiary condition imposed by Gauss' law on the physical states, namely
$Q|{\rm phys}>=0$. We will introduce in Section 3 a theory that 
generalizes Maxwell and in which
the $\sigma$ states become physical.

It will be interesting to verify the commutation rules of $\sigma$
with the vortex creation operator introduced in \cite{pol,nv,ap}, which is
given by
\be
\mu(x) \exp \lef \{ -ib \int_{T_{L(x)}} d^2 \xi F^{i0}(\xi,x^0) 
\del_i arg(\xi -x)\ri \}
\label{mu}
\ee
In this expression, $T_{L(x)}$ is the surface in Fig. 1, which 
consists of a plane of which a wedge containing the line $L(x)$, going from
$x$ to infinity has been taken out. This line corresponds to the 
cut of the $arg$ function appearing in (\ref{mu}). At the end we 
must take the cutoff $\delta \rightline 0$.
Again, the second part of (\ref{si}) commutes with $\mu$ and we 
immediately obtain
$$
\mu(x)\sigma(y) = \sigma(y) \mu(x) \exp \lef \{-i ab \int_{y,L(y)}^\infty
d\eta^i \int_{T_{L(x)}} d^2\xi \del_i^{(\eta)} arg(\eta -x)
\delta^2(\xi-\eta) \ri \}
$$
\be
\mu(x)\sigma(y) = \sigma(y) \mu(x) \exp \lef \{i ab\ \  arg(\vec y -\vec x)
\ri \}
\label{com2}
\ee
provided we choose the line $L(y)$ to belong to $T_{L(x)}$. This is precisely
the correct dual algebra satisfied by the operator $\mu$ with charged
operators \cite{vor,nv,ap}.

\subsection{ Locality}

Let us demonstrate now on general grounds, the line independence (locality)
of (\ref{cf1}). For this, let us perform the following change of 
functional integration variable in (\ref{cf1})
$$
A_\mu \rightarrow A_\mu + \Omega_\mu
$$
with
\be
\Omega_\mu = a \oint_{L'-L} d\xi_\mu \lef [\fr{1}{-\Box}\ri] (z- \xi) 
\label{tr}
\ee
where $L'$ is an arbitrary line going from $x$ to infinity. Under 
this transformation we have
\be
F_{\mu\nu} \rightarrow F_{\mu\nu}+ \C_{\mu\nu}(L'-L) - \C_{\nu\mu}(L'-L)
\label{tr1}
\ee
In Appendix A we show that under (\ref{tr}) and (\ref{tr1}) the exponent
in (\ref{cf1}), namely
\be
\cl[F_{\mu\nu}, \C_{\mu\nu}(L)] =
\fr{1}{4}
F_{\mu\nu}F^{\mu\nu} + F_{\mu\nu} \C^{\mu\nu}
+ \fr{1}{2} \C_{\mu\nu}\C^{\mu\nu}  
\label{ecf}
\ee
gets transformed as 
\be
\cl[F_{\mu\nu}, \C_{\mu\nu}(L)] \rightarrow \cl[F_{\mu\nu}, \C_{\mu\nu}(L')]
\label{tr2}
\ee
This shows that the correlation function $<\sigma(x)\sigma^\dagger(y)>$ is
independent of the curve $L$ appearing in the definition of the operator
$\sigma$, being therefore local.

\section{Correlation Functions}
\setcounter{equation}{0} 

\subsection{Maxwell Theory}

Let us explicitly evaluate in this Section the euclidean correlation 
functions involving the operator $\sigma$ in the case of pure Maxwell
theory. From (\ref{cf1}) we obtain, after performing the quadractic 
functional integral,
\be
<\sigma \sigma^\dagger> = \exp \lef\{ \fr{1}{2} \int d^3zd^3z' 
\C_{\mu\nu}(z)\C_{\alpha\beta}(z') F^{\mu\nu}_\sigma(z)
F^{\alpha\beta}_\lambda(z') D^{\sigma\lambda} (z-z') - S_L \ri\}
\label{cf2}
\ee
where 
\be
F^{\mu\nu}_\sigma(z) \equiv \del^\mu \delta^\nu\ _\sigma -
\del^\nu \delta^\mu\ _\sigma
\label{efe}
\ee
and $D^{\sigma\lambda}$ is the euclidean propagator of the Maxwell field,
given by
\be
D^{\sigma\lambda} = \lef (-\Box \delta^{\sigma\lambda}+ 
\lef ( 1 - \fr{1}{\xi} \ri )\del^\sigma\del^\lambda
\ri ) \lef [\fr{1}{(-\Box)^2}\ri]
\label{prop}
\ee
where $\xi$ is the gauge fixing parameter. $S_L$ in (\ref{cf2}) is the 
$\C_{\mu\nu}$ quadractic renormalization factor appearing in the last
term of (\ref{cf1}).
Inserting (\ref{efe}) and 
(\ref{prop}) in (\ref{cf2}), we see that only the gauge independent
first term of (\ref{prop}) contributes to it, thus confirming the gauge 
invariance of the $\sigma$ correlation function. Inserting $\C_{\mu\nu}$
in (\ref{cf2}) we get
$$
<\sigma \sigma^\dagger> = \exp \lef\{ \fr{a^2}{2} \sum_{i,j=1}^2
\lambda_i\lambda_j  \int_{x_i,L}^\infty d\xi^i \int_{x_j,L}^\infty d\eta^j
\times \ri. 
$$
\be
\lef.
\fr{\del^\nu_{(\xi)}}{-\Box} \fr{\del^\beta_{(\eta)}}{-\Box}
F^{i\nu}_\sigma(\xi) F^{j\beta}_\sigma(\eta)[-\Box]
\lef [\fr{1}{(-\Box)^2}\ri](\xi-\eta) -S_L \ri\}    
\label{cf3}
\ee
where $x_1 \equiv x, x_2 \equiv y, \lambda_1 \equiv +1, \lambda_2
\equiv -1$.
Using the identity 
\be
F^{i\nu}_\sigma(\xi) F^{j\beta}_\sigma(\eta) \equiv \epsilon^{i\nu\sigma}
\epsilon^{j\beta\lambda} \lef [ \del^\rho_{(\xi)}\del^\rho_{(\eta)}
\delta^{\sigma\lambda} - \del^\sigma_{(\xi)}\del^\lambda_{(\eta)} \ri]
\label{id}
\ee
in (\ref{cf3}) we see that the second term vanishes because of the
$\epsilon$'s and the contribution form the first term  
cancels the two $\fr{1}{\Box}$ pieces. Employing now the identity  
$\epsilon^{i\nu\sigma} \epsilon^{j\beta\sigma} \equiv 
\delta^{ij}\delta^{\nu\beta} - \delta^{i\beta}\delta^{\nu j} $, we 
see that the first term which is line dependent
is identically canceled by the renormalization factor $S_L$. The
second term is easily seen to be line independent, upon integration
over $\xi$ and $\eta$. The result is
\be
<\sigma \sigma^\dagger> = \exp \lef\{- \fr{a^2}{2} \sum_{i,j=1}^2
\lambda_i\lambda_j  F(x_i-x_j) \ri \}
\label{cf4}
\ee
where
\be
F(x) =\fr{1}{(-\Box)^2} \equiv \cf^{-1}\lef(\fr{1}{k^4}\ri) = 
\lim_{m\rightarrow \infty} \lef[\fr{1}{m} - \fr{|x|}{8\pi} \ri]
\label{f1}
\ee
Here $m$ is an infrared regulator used to define the inverse Fourier
transform of $\fr{1}{k^4}$. Introducing (\ref{f1}) in (\ref{cf4})
we see that the $m$-dependent part cancels and all infrared singularities
disappear. The remaining term gives
\be
<\sigma(x) \sigma^\dagger(y)> = \exp \lef\{-\fr{a^2}{8\pi} |x-y| \ri \}   
\label{cf5}
\ee
This is the final result for the $\sigma$ correlation function.  
The exponential decay would indicate a massive $|\sigma>$ 
state but as we saw this
is unphysical in pure Maxwell theory. The same expression as (\ref{cf5})
has been obtained in \cite{ap} for the gauge invariant part of the 
analogous operator. The present
formulation of charged operators, however, presents the advantage of
being gauge invariant and much simpler.

The correlation function of the vortex operator has been evaluated in
\cite{nv}, giving the result
\be
<\mu \mu^\dagger> = \exp \lef\{ \fr{\pi b^2}{|x-y|} 
 \ri \}
\label{cf6}
\ee
The large distance behavior of this indicates that $\mu$ does not create
genuine vortex states in this theory.

It will be instructive now to evaluate the mixed $\sigma-\mu$
correlation function
$$
<\sigma(x_1)\sigma^\dagger(x_2)\mu(y_1)\mu^\dagger(y_2)> =
Z^{-1} \int DA_\mu \expo 
\lef[ \fr{1}{4}
\lef (F_{\mu\nu}+ \B_{\mu\nu}\ri ) \lef (F^{\mu\nu}+ \B^{\mu\nu}\ri )
\ri.\ri.
$$
\be
\lef.\lef.
+ \lef(F_{\mu\nu}+ \B_{\mu\nu}\ri )\C^{\mu\nu}
+ \fr{1}{2} \C_{\mu\nu}\C^{\mu\nu}  \ri]\ri\}
\label{cf7}
\ee
The part corresponding to $\mu$ is associated with the $\B^{\mu\nu}$
external field, which is given by \cite{nv}
\be
\B^{\mu\nu}(z;x) = b\int_{T_{L(x)}}d^2\xi^{[\mu}\del^{\nu]}arg(\xi-x)
\delta^3(z-\xi)
\label{b}
\ee
where the surface $T_{L(x)}$ was defined above. Notice that a crossed 
$\B-\C$ renormalization term appeared in order ensure locality 
(path independence).
The above functional integral can be computed as before. The $\C^{\mu\nu}$-
part was just evaluated above. The $\B^{\mu\nu}$-part was calculated in 
\cite{nv}. The result is
\be
<\sigma(x_1)\sigma^\dagger(x_2)\mu(y_1)\mu^\dagger(y_2)> =  
\exp \lef\{- \fr{a^2}{8\pi} |x-y| + \fr{\pi b^2}{|x-y|} + CT \ri\}
\label{cf8}
\ee
where the crossed term is given by
$$
CT = iab \sum_{i,j=1}^2 \lambda_i \lambda_j \int_{x_i,L}^\infty d\xi^i
\int_{T_L} d^2 \eta_\mu \del_k arg(\xi -y_j)
\fr{\del^\rho_{(\xi)}}{-\Box} F^{i\rho}_\sigma(\xi) F^{\mu k}_\lambda(\eta)
D^{\sigma\lambda}(\xi-\eta) 
$$
\be
- \int d^3z \C_{\mu\nu} \B^{\mu\nu}
\label{cf9}
\ee
where the last piece is the $\B-\C$ term appearing in (\ref{cf7}).
Again only the gauge invariant first term of (\ref{prop}) contributes to 
(\ref{cf9}). Using the identity (\ref{id}), also here we find that the 
second term of (\ref{id})vanishes. 
The first term of (\ref{id}) contains two parts: one vanishes 
because $d\xi^i \bot d^2 \eta^j$, the other one gives
\be
CT = -iab \sum_{i,j=1}^2 \lambda_i \lambda_j 
\int_{T_L} d^2 \eta^\mu  arg(\eta - y_j)
\del_\mu \lef [\fr{1}{-\Box}\ri ](\eta - x_i) 
- \int d^3z \C_{\mu\nu} \B^{\mu\nu}
\label{cf10}
\ee
In Appendix B we show that 
\be
CT = A(x_i -y_j) \equiv n\ arg(x_i -y_j)
\label{ct}
\ee
where $n=0,\pm 1,\ldots$. We see that $CT$ is zero modulo a function
$arg(x_i -y_j)$. We have therefore
$$
<\sigma(x_1)\sigma^\dagger(x_2)\mu(y_1)\mu^\dagger(y_2)> =  
\exp \lef\{- \fr{a^2}{8\pi} |x_1-x_2| + \fr{\pi b^2}{|y_1-y_2|} +  
A(x_1 -y_1) + A(x_1 -y_2) \ri.
$$
\be
\lef.
+ A(x_2 -y_1) + A(x_2 -y_2) \ri\}
\label{cf11}
\ee
The ambiguities contained in the $A$-functions are nothing but a reflex
of the nontrivial commutation relations between $\sigma$'s and $\mu$'s
on the left hand side of (\ref{cf11}) \cite{odd,kc}. 
The various possible orderings of the operators
correspond to the different possibilities of $arg$-functions in the $A$'s.
This is a beautiful example, actually the first one in 2+1D, of how the 
functional integral can correctly describe the nontrivial algebra of
operators, even though the integration is performed over bosonic 
fields only.

\subsection{Nonlocal Maxwell Theory}

One can easily generalize the method for the evaluation of 
correlation functions of $\sigma$ and $\mu$ operators for the 
nonlocal generalization of the Maxwell theory given by
\be
\cl' = -\fr{1}{4} F^{\mu\nu}\lef[\fr{1}{(-\Box)^{1/2}}\ri ] F_{\mu\nu}
\label{mnl}
\ee
This theory was shown to describe the true electromagnetic interaction
of three-dimensional particles constrained to move on a plane \cite{qedpl}.
It also appeared in the bosonization of the free massless fermion in 
2+1 D \cite{bos}.

In this case, the $\sigma$ and $\mu$ operators would be described by
external fields analog to (\ref{c}) and (\ref{b}), given respectively by
\be
\C_\mu(z;x) = i a \int_{x,L}^\infty d\xi_\mu 
\lef[\fr{1}{(-\Box)^{1/2}}\ri] (z-\xi)
\label{c1}
\ee
and
\be
\B^{\mu\nu}(z;x) = b\int_{T(L)}d^2\xi^{[\mu}\del^{\nu]}arg(\xi-x)
\lef[\fr{1}{(-\Box)^{1/2}}\ri](z-\xi)
\label{b}
\ee
with the obvious modification of the renormalization factors. 

The quantization of theories described by lagrangians of the type of
(\ref{mnl}) has been performed in full detail in \cite{ru}. One can show
that $\sigma$ creates states carrying the charge $j^0 = \del_i \Pi^i$.
Again, however, the 0$^{th}$ component of the field equation corresponding
to (\ref{mnl}), namely $\del_i \Pi^i =0$ must be imposed as a constraint
over the physical states, implying that the $|\sigma>$ states
will also be unphysical here.

The evaluation of the correlation functions is entirely analogous to the 
case of pure Maxwell theory and we just show the results
\be
<\sigma(x) \sigma^\dagger(y)> = \fr{1}{ |x-y|^{a^2}}    
\label{cf55}
\ee
\be
<\mu(x) \mu^\dagger(y)> =  \fr{1}{|x-y|^{b^2}} 
\label{cf66}
\ee
and
$$
<\sigma(x_1)\sigma^\dagger(x_2)\mu(y_1)\mu^\dagger(y_2)> =  
\fr{1}{ |x_1-x_2|^{a^2}}  \fr{1}{ |y_1-y_2|^{b^2}}   
\exp \lef\{               
A(x_1 -y_1) + A(x_1 -y_2) + \ri.
$$
\be
\lef.
A(x_2 -y_1) + A(x_2 -y_2) \ri\}
\label{cf11}
\ee
Notice that $a$ and $b$ are dimensionless now. Again we have the 
$A(x-y)$ factors in the mixed function, indicating that the $\sigma$
and $\mu$ operators will have the same commutation relations as in the 
Maxwell theory. This confirms the result of the quantization method
developed in \cite{ru}.

\section{The Dual Theory}
\setcounter{equation}{0}

We saw in Section 2 that the charged states created by $\sigma$ were
unphysical in pure Maxwell theory. We will introduce now a generalization
of this theory in which these charged $\sigma$-states do exist in the 
physical space. 

First of all, let us introduce a vector gauge field $W_\mu$, in 
terms of which we can express the vacuum functional of Maxwell theory as
the following euclidean functional integral
$$
Z_A = \int DA_\mu \exp \lef \{ - \int d^3z \fr{1}{4} F^{\mu\nu}F_{\mu\nu}
\ri \} =
$$
\be
= Z^{-1}_0 \int DW_\mu \exp \lef \{ - \int d^3z \lef [ 
\fr{1}{4} W^{\mu\nu} \lef[\fr{1}{-\Box} \ri] W_{\mu\nu} 
+ i \epsilon^{\mu\alpha\beta} 
W_\mu\del_\alpha A_\beta \ri ] + \cl_{GF} \ri \}
\label{zs}
\ee
The field equation associated to the lagrangian of the $W_\mu-A_\mu$ theory
\be
\cl_{WA} =\fr{1}{4} W^{\mu\nu} \lef[\fr{1}{-\Box} \ri] W_{\mu\nu}
- \epsilon^{\mu\alpha\beta}W_\mu \del_\alpha A_\beta
\label{wa}
\ee
can be written as
\be
\del_\nu F^{\mu\nu} = \epsilon^{\mu\alpha\beta} \del_\alpha W_\beta
\label{feq}
\ee
We see that the topological current of the $W_\mu$-field, namely
$J^\mu = \epsilon^{\mu\alpha\beta} \del_\alpha W_\beta $ becames the 
source of the electromagnetic field, that is, an electric current. 
If we integrate over $A_\mu$ in
(\ref{zs}), however, we get
\be
Z_A =
Z^{-1}_0 \int DW_\mu \exp \lef \{ - \int d^3z \lef [
\fr{1}{4} W^{\mu\nu} \lef[\fr{1}{-\Box} \ri] W_{\mu\nu} 
+ \cl_{GF}\ri ] \ri \} \delta [\epsilon^{\mu\alpha\beta} 
\del_\alpha W_\beta]
\label{z1}
\ee
We conclude that pure Maxwell theory can be described by the pure $W_\mu$
theory with the constraint that we should be in the sector where the 
topological charge (electric charge) is identically equal to zero.

Let us consider now the theory given by
\be
\cl_W = - \fr{1}{4} W^{\mu\nu} \lef[\fr{1}{-\Box} \ri] W_{\mu\nu} =
\fr{1}{2} J^\mu \lef[\fr{1}{-\Box} \ri] J_\mu
\label{lw}
\ee
where the constraint $J^\mu \equiv 0$ has been relaxed. Of course this is
no longer Maxwell theory, but a generalization of it where  
we can have physical
states with a nonzero topological charge. As we can infer from the 
second part of (\ref{lw}) the interaction of these charges will be identical
to that of usual charges interacting through the electromagnetic field.

Let us introduce now the operator $\tilde\sigma$ which is going to create
the states carrying topological charge (electric charge) 
in the theory described by $\cl_W$. Consider
$$
\tilde\sigma(x) = \exp \lef \{i \fr{a}{2} \int_{x,L}^\infty d\xi_\lambda
\epsilon^{\lambda\mu\nu}\lef[ \fr{W^{\mu\nu}}{-\Box}\ri ] \ri \}
$$
\be
\tilde\sigma(x) = \exp \lef \{i a \int_{x,L}^\infty d\xi^i 
\epsilon^{ij} \Pi^j \ri \}
\label{sit}
\ee
where $\Pi^i = \fr{W^{0i}}{-\Box}$. 
Observe that if we use the field equation (\ref{feq}) in (\ref{sit})
we can see that $\tilde\sigma = \sigma$, as given by (\ref{si}).

The quantization of theories of the 
type described by (\ref{lw}) has been carefully studied in \cite{ru}.
One can show that 
\be
[J^0(x), \tilde\sigma (y)] = a \tilde\sigma (y) \delta^2(x-y)
\label{cj}
\ee
indicating that indeed $\tilde\sigma$ does create states with $a$ units
of charge. 

Let us consider now the correlation function of $\tilde\sigma$. This
can be written as
\be
<\tilde\sigma (x)\tilde\sigma^\dagger(y)> = Z_0^{-1} \int DW_\mu
\exp \lef \{ - \int d^3z \lef [ 
\fr{1}{4}( W^{\mu\nu}+ \bb^{\mu\nu} ) \lef[\fr{1}{-\Box} \ri] 
( W_{\mu\nu} + \bb_{\mu\nu} ) \ri ] \ri \} 
\label{cft1}
\ee
where $\bb^{\mu\nu}= \bb^{\mu\nu} (z;x)-\bb^{\mu\nu} (z;y)$ with
\be
\bb^{\mu\nu} (z;x) = a \int_{x,L}^\infty d\xi_\lambda 
\epsilon^{\lambda\mu\nu}
\delta^3 ( z-\xi)
\label{b1}
\ee
The crossed $W-\bb$ term is easily identified with the $\tilde\sigma$ 
operators (\ref{sit}). The $\bb-\bb$ term, as before is a renormalization 
factor which renders (\ref{cft1}) path independent. The locality of 
(\ref{cft1}) can be demonstrated in general by making a transformation
similar to (\ref{tr}) but with $\Omega_\mu$ replaced by \cite{nv}
\be
\tilde\Omega_\mu = a \int_{S(L'-L)} d\xi_\mu \delta^3(z-\xi)
\label{ot}
\ee
where $S(L'-L)$ is an arbitrary surface bounded by $L$ and a general
curve $L'$. 

In order to evaluate $<\tilde\sigma\tilde\sigma^\dagger>$ 
we insert (\ref{b1}) in (\ref{cft1}) and integrate over $W_\mu$, obtaining
$$
<\tilde\sigma (x)\tilde\sigma^\dagger(y)>= \exp \lef \{ \fr{a^2}{2}
\sum_{i,j=1}^2 \lambda_i\lambda_j \int_{x_i,L} d\xi^i \int_{x_j,L} d\eta^j
\times \ri. 
$$
\be
\lef.   
\epsilon^{i\mu\alpha} \fr{\del_\mu^{(\xi)}}{-\Box}
\epsilon^{j\nu\beta} \fr{\del_\nu^{(\eta)}}{-\Box}
[-\Box \delta^{\alpha\beta} +(1-\fr{1}{\xi}) \del^\alpha\del^\beta ]
\lef [ \fr{1}{-\Box} \ri ] - S_L   \ri \}
\label{cft2}
\ee
In this expression, $S_L$ is the $\bb-\bb$ term of (\ref{cft1}) and the 
expression between brackets is the euclidean propagator of the $W_\mu$
field, corresponding to (\ref{lw}). We use the same convention for the 
$\lambda$'s as before. Only the gauge invariant first part of the
propagator contributes in (\ref{cft2}). The rest identically vanishes.
Using the identity for the $\epsilon$'s appearing after (\ref{id}) we
get two terms. One of them is path dependent and is identically canceled
by $S_L$. The remaining term is path independent and gives an expression 
identical to (\ref{cf4}). We therefore obtain 
\be
<\tilde\sigma(x) \tilde\sigma^\dagger(y)> = 
\exp \lef\{-\fr{a^2}{8\pi} |x-y| \ri \}   
\label{cft3}
\ee
The massive $|\sigma>$ states are no longer forbidden by the condition
to be imposed on the physical subspace. We should remark, however that
they will not be asymptotic, because of the short distance behavior of
the correlation function (\ref{cft3}).

It is interesting to observe that also in 3+1D a theory dual to 
electrodynamics was found \cite{qed} in complete analogy to what
we did here. This theory is formulated 
in terms of an antisymmetric tensor gauge field. In the generalization
where the zero topological charge constraint is relaxed we find massless
charged states which can be asymptotic for a certain value of the charge
\cite{qed}.

\section{Final Comments}           

We have seen how to construct creation operators of charged states,
which are 
expressed only in terms of the gauge field, in three dimensional
theories whose lagrangian is quadractic in the field intensity tensor.
This charge bearing gauge invariant operators are always dual to the
magnetic flux carrying operator (vortex operator) 
in each of the theories considered. Only in the theory introduced in 
Section 3, however, which generalizes the simple Maxwell case, 
by the introduction of nontrivial topological sectors, these
charged states are physical. This theory is completely analogous to
the one investigated in \cite{qed} in connection to four-dimensional QED.
Even in the cases where the states are unphysical, however, they are
interesting and worth to be investigated, for the following reason.
Putting together the charged operators introduced here and the magnetic
flux operators introduced in \cite{pol,vor,nv} we have a complete picture
of duality in 2+1D. This has the same structure as the one found in 1+1D,
involving scalar fields \cite{cm,odd}. The novelty here is that since we 
have gauge theories underlying the construction, the states created by the 
corresponding operators are not always physical. It is known \cite{odd}
that in 1+1D, this structure is responsible for the bosonization of
fermions hence it is quite natural to expect that this is also true in
higher dimensions. The case of a free 
massless fermion field in 2+1D has been
considered in \cite{bos}. Some general results have been obtained more
recently in \cite{mar}. We are presently investigating the application 
of the dual operators we found in 2+1D in connection with the method 
of bosonization.

\appendix

\renewcommand{\thesection}{\Alph{section})}

\renewcommand{\theequation}{\Alph{section}.\arabic{equation}}

\section{ Appendix A}
\setcounter{equation}{0}

Let us demonstrate here Eq.(\ref{tr2}) and thereby establish the path 
independence of (\ref{cf1}). Inserting (\ref{tr1}) in (\ref{ecf}) 
we get
\be
\cl[F_{\mu\nu}, \C_{\mu\nu}(L)] \rightarrow \cl[F_{\mu\nu}, \C_{\mu\nu}(L)]
+ \Delta \cl[F_{\mu\nu}, \C_{\mu\nu}(L)]
\label{a1}
\ee
where
$$
\Delta \cl[F_{\mu\nu}, \C_{\mu\nu}(L)] = \fr{1}{2} \C_{\mu\nu}\C_{\mu\nu}
(LL+L'L'-2LL') -\fr{1}{2} \C_{\mu\nu}\C_{\nu\mu}(LL+L'L'-2LL')
$$
\be
+\C_{\mu\nu}(L)\C_{\mu\nu}(L'-L)-\C_{\mu\nu}(L)\C_{\nu\mu}(L'-L)+
+\C_{\mu\nu}(L'-L) F^{\mu\nu}
\label{a2}
\ee
In the above expression we have schematically represented the $L$ and $L'$
dependence in an obvious way. The crossed $LL'$ terms cancel. 
The $LL$ and $L'L'$ $\mu\nu-\nu\mu$ terms are of the form
$$
\fr{1}{2}\C_{\mu\nu}(L)\C_{\nu\mu}(L)- (L \leftrightarrow L')=
$$
\be
= \int_{z,L}^\infty d\xi_\mu \int_{z',L}^\infty d\eta_\nu 
\del_{(\xi)}^\mu\del_{(\eta)}^\nu F(\xi-\eta) - (L \leftrightarrow L') =0
\label{a3}
\ee
This is zero because the above integral is clearly independent of the 
curve $L$. Now, inserting the remaining terms of (\ref{a2}) in (\ref{a1})
we see that
the $\C_{\mu\nu}F^{\mu\nu}$ and  $\C_{\mu\nu}\C^{\mu\nu}$ terms
in (\ref{a2}) will combine with the corresponding ones in (\ref{ecf})
thereby exchanging $L$ for $L'$ in $\cl[F_{\mu\nu}, \C_{\mu\nu}(L)]$.
This establishes Eq.(\ref{tr2}).

\section{ Appendix B}
\setcounter{equation}{0}

Let us demonstrate here Eq.(\ref{ct}). According to Fig. 2, the first
part of (\ref{ct}) is given by
$$
I = -iab \int_{T_{L(y_j)}} d^2 \xi^\mu  arg(\xi - y_j)
\del_\mu \lef [\fr{1}{4\pi|\xi - x_i|} \ri ]    = 
$$
$$
= -iab
\fr{H}{4\pi} \int_0^\infty dr r \int_0^{2\pi} d\theta 
\fr{\theta}{[D^2 + r^2 -2Lr \cos(\theta -\alpha)]^{3/2}} =
$$
\be
=-iab \fr{H}{4\pi} \int_0^\infty dr r \int_{-\alpha}^{2\pi-\alpha} d\omega 
\fr{\omega + \alpha}{[D^2 + r^2 -2Lr \cos\omega]^{3/2}} 
\label{b1}
\ee
Now,
$$
I' = \int_{T_{L(y_j)}} d^2 \xi^\mu 
\del_\mu\lef[\fr{1}{4\pi|\xi - x_i|}\ri] = \int_{T_{L(y_j)}} 
\fr{ {\vec {dS}} \cdot {\hat R}}{4\pi R^2} =
\fr{H}{4\pi} \int_0^\infty dr r \int_{-\alpha}^{2\pi-\alpha} d\omega 
\fr{1}{[D^2 + r^2 -2Lr \cos\omega]^{3/2}} 
$$
\be
= \fr{\Omega(x_i;T_{L(y_j)})}{4\pi}
\label{b2}
\ee
where $\Omega(x_i;T_{L(y_j)})$ is the solid angle formed by $x_i$ 
and the surface $T_{L(y_j)}$. This is defined up to $4\pi$ factors. Actually,
in the limit when we take out the cutoff from $T_{L(y_j)}$ we have
$\Omega(x_i;T_{L(y_j)}) = 2\pi + n 4\pi$ where $n$ is an arbitrary integer.

Comparing (\ref{b2}) and (\ref{b1}) we see that
\be
I=-iab\  arg(x_i -y_j) \fr{\Omega(x_i;T_{L(y_j)})}{4\pi} 
-iab \fr{H}{4\pi} \int_0^\infty dr r \int_{-\alpha}^{2\pi-\alpha} d\omega 
\fr{\omega}{[D^2 + r^2 -2Lr \cos\omega]^{3/2}} 
\label{b3}
\ee

Using now (\ref{b}) and (\ref{c}), we see that the second part of 
(\ref{ct}) is precisely $-I$. Hence (\ref{b3}) is canceled and in particular
the line(cut)-dependent second piece. Since the first term of (\ref{b3}),
however, is ambiguously defined we can always choose the $n$-integers
in the $\Omega(x_i;T_{L(y_j)})$'s in such a way 
that there always remains a factor
$A(x_i -y_j) \equiv n\ arg(x_i -y_j)$
for arbitrary $n$. This proves (\ref{ct}). Each prescription for $n$
corresponds to a particular ordering of operators on the left hand side
of (\ref{cf11}).

\end{document}